\documentclass[twocolumn,aps,showpacs,prl]{revtex4}
\usepackage{graphicx}
\usepackage{epstopdf}
\usepackage{amsmath}
\usepackage{multirow}

\begin{document}

\title{Atomic and electronic structures of ternary iron arsenides
$A$Fe$_2$As$_2$(001) surfaces ($A$=Ba, Sr, or Ca)}
\author{Miao Gao$^{1}$}
\author{Fengjie Ma$^{1,2}$}
\author{Zhong-Yi Lu$^{1}$}\email{zlu@ruc.edu.cn}
\author{Tao Xiang$^{3,2}$}\email{txiang@aphy.iphy.ac.cn}

\date{\today}

\affiliation{$^{1}$Department of Physics, Renmin University of
China, Beijing 100872, China}

\affiliation{$^{2}$Institute of Theoretical Physics, Chinese Academy
of Sciences, Beijing 100190, China }

\affiliation{$^{3}$Institute of Physics, Chinese Academy of
Sciences, Beijing 100190, China }

\begin{abstract}

By the first-principles electronic structure calculations, we find
that energetically the most favorable cleaved $A$Fe$_2$As$_2$(001)
surface ($A$=Ba, Sr, or Ca) is $A$-terminated with a
$(\sqrt{2}\times \sqrt{2})R45^{\circ}$ or $(1\times 2)$ order. The
$(1\times 2)$ ordered structure yields a $(1\times 2)$ dimerized STM
image, in agreement with the experimental observation. The $A$ atoms
are found to diffuse on the surface with a small energy barrier so
that the cleaving process may destroy the $A$ atoms ordering. At the
very low temperatures this may result in an As-terminated surface
with the $A$ atoms in randomly assembling. The As-terminated
BaFe$_2$As$_2$ surface in orthorhombic phase is $(\sqrt{2}\times
\sqrt{2})R45^{\circ}$ buckled, giving rise to a switchable
$(\sqrt{2}\times \sqrt{2})R45^{\circ}$ STM pattern upon varying the
applied bias. No any reconstruction is found for the other
As-terminated surfaces. There are surface states crossing or nearby
the Fermi energy in the As-terminated and $(1\times 2)$
$A$-terminated surfaces. A unified physical picture is thus
established to help understand the cleaved $A$Fe$_2$As$_2$(001)
surfaces.

\end{abstract}


\maketitle


Recently great interest has been devoted to the investigation of
superconduction properties of layered iron pnictides upon doping or
high pressure\cite{kamihara}. Both the neutron scattering and
theoretical studies revealed that the undoped parent compounds at
atmospheric pressure are in either collinear or bi-collinear
antiferromagnetic order below either tetragonal-orthorhombic or
tetragonal-triclinic structural transition temperature
\cite{cruz,dong,ma,bao}. The surfaces of high-quality single crystal
Ba(Sr)Fe$_2$As$_2$ have also been probed by Scanning Tunneling
Microscopy (STM) \cite{shpan,zqwang,jghou}. However, the
contradictory observations on the nature of terminated surfaces have
been found. On a single crystal of BaFe$_2$As$_2$ cleaved at 20K,
the STM images observed at 4.3K showed a dominant ($\sqrt{2}\times
\sqrt{2})R45^{\circ}$ (denoted by $\sqrt{2}$ thereafter) order,
which was interpreted as due to the ordering of As atoms on the
terminated surface\cite{shpan}. The STM measurement on another
BaFe$_2$As$_2$ single crystal cleaved at 120K also revealed a
$\sqrt{2}$ order, but it was interpreted as due to the ordering of
Ba atoms on the terminated surface\cite{jghou}. In contrast, the STM
measurement observes both $\sqrt{2}$ and $(1\times 2)$ orders on a
SrFe$_2$As$_2$ single crystal\cite{zqwang}. These reflect there will
be rich surface properties distinct from the bulk. A thorough
investigation on the underlying atomic structures and electronic
structures of the cleaved surfaces is important to resolve these
contradictions. This is also important to the understanding of
experimental data measured by Angle-Resolved Photoemission
Spectroscopy (ARPES) which is a surface-sensitive probe.

In this Letter, we present the results of first-principles
electronic structure calculations on the (001) surfaces of
$A$Fe$_2$As$_2$ (denoted as $A$122 thereafter)($A$=Ba, Sr, or Ca).
We find that the $A$-terminated surface with either $\sqrt{2}$ or
$(1\times 2)$ order is energetically most favorable in these
materials. Cleaving at very low temperature may lift $A$ atoms free
in a large area, resulting in an ordered As layer to expose on the
surface. Correspondingly, there are the surface electronic states
induced crossing or nearby the Fermi energy. These surfaces lead to
different STM patterns and affect strongly low energy electronic
structures of the compounds.

In our calculations the plane wave basis method was used
\cite{electron,pwscf}. We adopted the generalized gradient
approximation of Perdew-Burke-Ernzerhof \cite{pbe} for the
exchange-correlation potentials. The ultrasoft pseudopotentials
\cite{vanderbilt} were used to model the electron-ion interactions.
After the full convergence test, the kinetic energy cut-off and the
charge density cut-off of the plane wave basis were chosen to be 408
eV and 4082 eV, respectively. The Gaussian broadening technique was
used and a mesh of $16\times 16$ k-points were sampled for the
$(1\times 1)$ irreducible surface Brillouin-zone integration. We
modeled the surface using a periodically repeated slab of six FeAs
layers and five or seven $A$ layers plus a vacuum layer of 20 to 25
\AA\ with inversion symmetry through the center of the slab. We
froze the two innermost middle FeAs layers with one $A$ layer and
allowed all other atoms to relax.

In this Letter we call a surface unit cell in reference to a square
lattice with $\sim$4\AA\ periodicity since both As atoms and $A$
atoms sit at such a square lattice in the bulk. A (1$\times$1) unit
cell is thus a 4\AA$\times$4\AA\ square. After cleaving, the surface
of $A$Fe$_2$As$_2$ can terminate either at an $A$-atom layer or at
an As-atom layer. An Fe-atom layer will not expose on the surface
due to the strong chemical bonding between Fe and As ions. On
average, half of $A$ atoms should remain on the surface to balance
the chemical valence. These $A$ atoms can only form either
$\sqrt{2}$ or $(1\times 2)$ order, derived by half of $A$ atoms
being removed alternatively from the original (1$\times$1) square
lattice of $A$ atoms along the square-side or square-diagonal
direction. In reality, there always yield two surfaces after
cleaving. It is expected that the cleaved surface would be
$A$-terminated with a $\sqrt{2}$ or $(1\times 2)$ order. On the
other hand, if one of cleaved surfaces is As-terminated, the other
will be $A$-terminated with all the $A$ atoms remaining at $(1\times
1)$ lattice.

\begin{table}
\caption{$A$Fe$_2$As$_2$ (001) surface energy differences ($A$=Ba,
Sr, or Ca) between the combinations of the two
$(\sqrt{2}\times\sqrt{2})R45^{\circ}$ $A$-terminated surfaces
(denoted as 2x$(\sqrt{2})$), the two $(1\times 2)$ $A$-terminated
surfaces (denoted as 2x(1x2)), and the $(1\times 1)$ $A$-terminated
and $(1\times 1)$ As-terminated surfaces (denoted 2x(1x1)) in
nonmagnetic tetragonal phase (tetragonal) and antiferromagnetic
orthorhombic phase (orthorhombic), respectively. Here we set the
surface energy of (2x(1x1)) to zero. The unit is meV/($1\times 1$
cell), whereas $1\times 1$ cell is a 4\AA$\times$4\AA\ square.
}\label{cleave}
\begin{tabular}{|c|c|c|c|c|c|c|}
  \hline
  $A$Fe$_2$As$_2$ & \multicolumn{3}{c|}{tetragonal}
  &\multicolumn{3}{c|}{orthorhombic } \\
  \cline{2-7}
    (001) & 2x$(\sqrt{2})$&2x(1x2)&2x(1x1)&
   2x$(\sqrt{2})$&2x(1x2)&2x(1x1) \\
  \cline{1-7}
  BaFe$_2$As$_2$ &-254.8 & -151.0 & 0.0 & -325.0 &-221.8 & 0.0\\
  \cline{1-7}
  SrFe$_2$As$_2$& -181.9 &-97.4 & 0.0 & -261.9 & -194.7 & 0.0 \\
  \cline{1-7}
  CaFe$_2$As$_2$ &-101.0& -107.8& 0.0 & -224.8& -210.1 & 0.0  \\
  \hline
\end{tabular}
\end{table}

\begin{figure}
\includegraphics[height=17.5cm]{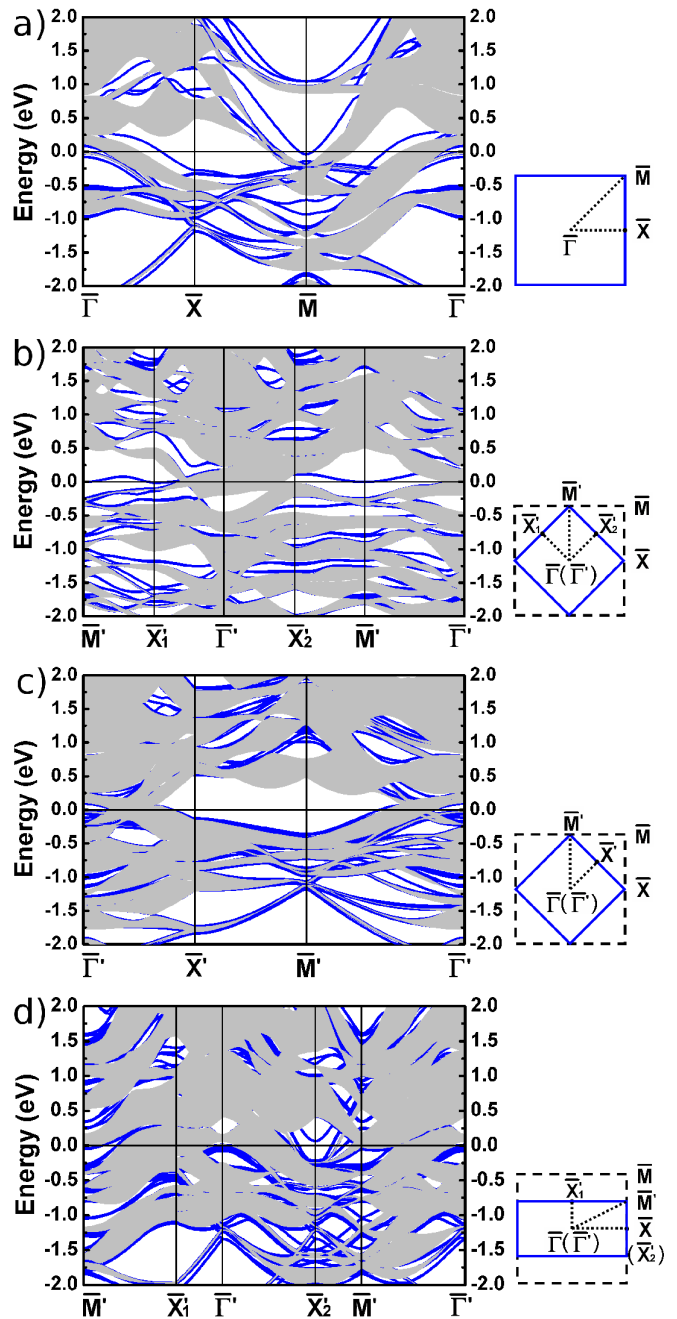}
\caption{(Color online) The calculated surface electronic band
structure of $A$Fe$_2$As$_2$ ($A$=Ba, Sr, or Ca) (001) surface
reported along the high-symmetry lines of the irreducible surface
Brillouin zone. Fermi level sets to zero. Shaded areas correspond to
the surface-projected bulk states while thicker blue lines
correspond to surface states. The surface Brillouin zones are given
next to the surface band structures whereas the dotted line squares
are $(1\times 1)$ surface Brillouin zone. (a) (1$\times $1)
As-terminated BaFe$_2$As$_2$ surface in the tetragonal phase; (b)
($\sqrt{2}\times\sqrt{2})R45^{\circ}$ As-terminated BaFe$_2$As$_2$
surface in the orthorhombic phase; (c)
($\sqrt{2}\times\sqrt{2})R45^{\circ}$ Ba-terminated BaFe$_2$As$_2$
surface in the orthorhombic phase; (d) ($1\times 2$) Ca-terminated
CaFe$_2$As$_2$ surface in the tetragonal phase. Note that ($1\times
1$) surface Brillouin zone is folded in (b), (c), and (d). }
\label{fig1}
\end{figure}

In order to clarify the cleaved surface structure, we calculated the
$A$-terminated surfaces with $(1\times 1)$, $\sqrt{2}$, and
$(1\times 2)$ orders described above and the As-terminated surface
with $(1\times 1)$ order, respectively. Table \ref{cleave}
summarizes the numerical results. We find that energetically the
most favorable cleaved $A$122 (001) surface is $A$-terminated. For
Ba122 it is $\sqrt{2}$-ordered while for Ca122 (Sr122) it is mainly
$(1\times 2)$ ($\sqrt{2}$)-ordered with partly $\sqrt{2}$ ($(1\times
2)$) order coexisting as the energy difference between the $(1\times
2)$ and $\sqrt{2}$ orders becomes small. This may be due to the
electropositivity becoming smaller in the line of Ba, Sr, and Ca.

Realistically, the kinetics of cleaving process plays an important
role in the formation of the cleaved surfaces of a sample. To
understand this more clearly, we carried out a calculation to
simulate a cleaving process. We divided a sample into two parts by
separating two neighbor FeAs layers adiabatically away from each
other. The $A$ layer between the two neighbor FeAs layers is either
divided half by half into each of the FeAs layers or kept as a whole
in one of FeAs layers in the separation process. It turns out that
no significant separation barrier energies are found in these two
cases. This suggests that the energetics shown in Table \ref{cleave}
may well describe the cleaved surfaces.

We further studied the diffusion of an $A$ atom on the As-terminated
surface. The energy barrier to move an $A$ atom from one equilibrium
position to another is found to be small so that the fast cleaving
process may destroy the $A$ atoms ordering. In particular, a very
low temperature cleaving can yield a metastable As-terminated
surface with randomly assembled $A$ atoms. The low energy electronic
structure is expected to be strongly modified on the As-terminated
surface since the unbalanced chemical valence in the surface FeAs
layer will induce surface states crossing or nearby the Fermi
energy. This may further induce surface reconstruction through
buckling or dimerization of the surface As atoms.

In the tetragonal phase, we don't find any reconstruction like
buckling or dimerization on the As-terminated $A$122 surfaces except
small inwards relaxation. Nevertheless, a number of surface states
are induced within the bulk energy gap regions, especially those
partially filled surface states nearby the Fermi energy (see Fig.
\ref{fig1}(a)).

\begin{figure}
\includegraphics[width=8.6cm]{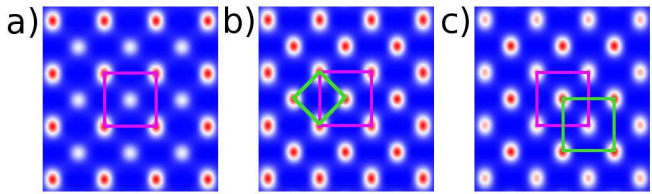}
\caption{(Color online) Simulated STM images of the
$(\sqrt{2}\times\sqrt{2})R45^{\circ}$-ordered As-terminated
BaFe$_2$As$_2$ (001) surface in orthorhombic phase with bias
voltages of (a) 30meV, (b) -30meV, and (c) -100meV. The large
squares indicate $(\sqrt{2}\times\sqrt{2})R45^{\circ}$ surface unit
cells, while the small square indicates $(1\times 1)$ surface unit
cell. Note that the simulated STM images for As-terminated
Sr(Ca)Fe$_2$As$_2$ surface show the $(1\times 1)$ order, similar to
(b). } \label{fig2}
\end{figure}

In the orthorhombic phase, the As-terminated $A$122 surface takes a
$\sqrt{2}$ magnetic surface unit cell. Our calculations show that
there is neither buckling nor dimerization except in Ba122 where a
small buckling of $\sim$0.1\AA\ occurs with an energy gain of about
3.9 meV/($1\times 1$ cell). Such a small buckling can nevertheless
affect strongly the STM measurement, as shown in Fig.~\ref{fig2}.
For a positive bias or a small negative bias (Fig. \ref{fig2}a), the
calculated STM image shows a $\sqrt{2}$ order, in which the buckled
inward As atoms show strong bright spots while the buckled outward
As atoms are (nearly) invisible. For a mediate bias between -30 meV
and -50 meV (Fig. \ref{fig2}b), a $(1\times 1)$ order of bright
buckled As is found, in which all the buckled As atoms show the
bright spots. For a large negative bias below -50 meV (Fig.
\ref{fig2}c), a $\sqrt{2}$ order, in which the buckled inward As
atoms are nearly invisible while the buckled outward As atoms are
brightest, is found. Here a positive (negative) bias means the STM
probes the states above (below) the Fermi energy. These results show
that the STM image can be switched from a $\sqrt{2}$ pattern shifted
by $\sim$4\AA\ to another $\sqrt{2}$ pattern through a $(1\times 1)$
order by varying the bias voltage. For the As-terminated surface in
Sr(Ca)Fe$_2$As$_2$, the calculated STM images show a $(1\times 1)$
order, similar to the one shown in Fig. \ref{fig2}(b).

The surface band structure (Fig. \ref{fig1}(b)) shows that there are
a number of surface states induced within the bulk energy gap
regions in the As-terminated Ba122. Here we remind that the
nonmagnetic tetragonal surface Brillouin zone is reduced by half
through folding in the antiferromagnetic orthorhombic surface
Brillouin zone. As shown in Fig. \ref{fig1} (b), the tetragonal
$\bar{M}$ and $\bar{\Gamma}$ become $\bar{\Gamma}^{\prime}$ in the
folded Brillouin zone, and the tetragonal $\bar{X}$ is now located
at $\bar{M}^{\prime}$.

The buckling of a surface is to gain energy by lifting the
degeneracy of energy band due to breaking of mirror symmetry. The
buckling of pairs of As atoms at the surface causes the surface
atoms to partially dehybridize. This causes some electronic charge
transfer from the inward atoms to the outward atoms so that the
initial partially filled surface states become insulating (see Fig.
\ref{fig1}(b)). This is the reason why no buckling happens in the
nonmagnetic tetragonal phase whereas the energy gain by the charge
transfer is smaller due to the stronger metallicity. This is also
the reason why the STM can observes the inward and outward atoms at
a positive and negative bias, respectively. On the other hand, the
buckling will cause a loss of the surface lattice distortion energy.
These competing effect determines whether or not the buckling will
take place. In contrast, for Sr(Ca)122, no reconstruction is found
in the As-terminated surface. This is because in Sr(Ca)122 the
lattice constants are smaller than that in Ba122 and that the
lattice distortion will cost more energy than the buckling energy
gain. This is similar to the buckled Si(001) surface versus the
non-buckled C(001) surface. As a result, the As-terminated Sr(Ca)122
surfaces always yield a $(1\times 1)$ order in either tetragonal or
orthorhombic phase, similar to Fig. \ref{fig2}(b).

\begin{figure}
\includegraphics[width=8.0cm]{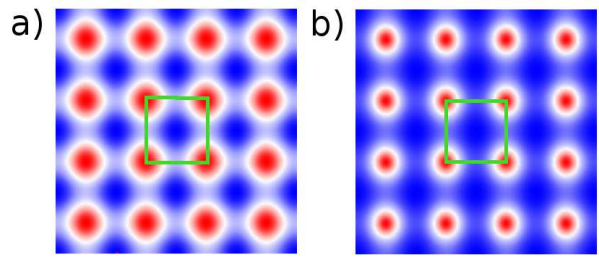}
\caption{(Color online) Simulated STM images of the
$(\sqrt{2}\times\sqrt{2})R45^{\circ}$ ordered Ba-terminated
BaFe$_2$As$_2$ (001) surface in orthorhombic phase with bias
voltages of (a) 100meV and (b) -100meV. The squares indicate
$(\sqrt{2}\times\sqrt{2})R45^{\circ}$ surface unit cells. Note that
the similar simulated STM images are obtained for the
$(\sqrt{2}\times\sqrt{2})R45^{\circ}$ ordered Sr(Ca)-terminated
Sr(Ca)Fe$_2$As$_2$ surface.} \label{fig3}
\end{figure}

For the $\sqrt{2}$-$A$-terminated $A$122 surface, no lattice
reconstruction (buckling or dimerization) is found. One expects that
those surface states crossing the Fermi energy in the As-terminated
surfaces (see Fig. \ref{fig1} (a) and (b)), induced by the
unbalanced chemical valence, would be eliminated by the $A$ atoms
donating charge into the FeAs layer to make the chemical valence
balanced in the whole surface layer. This is verified by our
calculations, as shown in Fig. \ref{fig1} (c). The remaining surface
states are mainly distributed around the edges of the bulk energy
gap regions, induced by the reduced surface symmetry. The calculated
STM images show a $\sqrt{2}$-ordering pattern in both tetragonal and
orthorhombic phases (see Fig.\ref{fig3}), in agreement with the
recent STM observation\cite{jghou}.

\begin{figure}
\includegraphics[width=8.0cm]{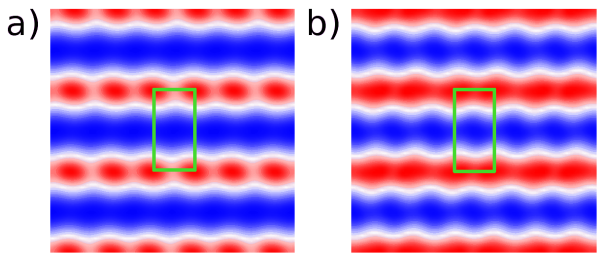}
\caption{(Color online) Simulated STM images of the $(1\times 2)$
ordered Ca-terminated CaFe$_2$As$_2$ (001) surface in orthorhombic
phase with bias voltages of (a) 100meV and (b) -100meV. The squares
indicate $(1\times 2)$ surface unit cells. Note that the similar
simulated STM images are obtained for the $(1\times 2)$ ordered
Sr(Ba)-terminated Sr(Ba)Fe$_2$As$_2$ surface. } \label{fig4}
\end{figure}

For the $(1\times 2)$-$A$-terminated $A$122 surfaces, no any other
reconstruction (buckling or dimerization) is found for the surface
$A$ atoms either. However, in both tetragonal and orthorhombic
phases, the surface $A$ atoms will move downwards so that the
underneath As atoms move closely with each other by $\sim$0.05 \AA.
This small dimerization is perpendicular to the surface $A$ atom row
and located in the troughs between the $A$ atom rows, and lead to
some new surface states near the Fermi energy (Fig. \ref{fig1}(d)).
However, the calculated STM images show the apparent dimer rows
along the surface $A$ atom rows rather than the troughs, due to the
hybridization between the surface $A$ atoms and As atoms (Fig.
\ref{fig4}), also in agreement with the recent experimental STM
observation\cite{zqwang}.

In summary, a unified physical picture on cleaved $A$Fe$_2$As$_2$
(001) surfaces ($A$=Ba, Sr, or Ca) is established based on the first
principles band structure calculations. The energetically most
favorable cleaved surface is $A$-terminated with $(\sqrt{2}\times
\sqrt{2})R45^{\circ}$ or $(1\times 2)$ order. For Ba122, the surface
$A$ atoms are found to be in the $(\sqrt{2}\times
\sqrt{2})R45^{\circ}$ order, in agreement with the experimental
observation\cite{jghou}. However, as there is a small diffuse energy
barrier for $A$ atoms, a fast cleaving at very low temperature may
yield an As-terminated ordered surface with randomly assembled $A$
atoms as a metastable surface structure. This As-terminated surface
is also found to be $(\sqrt{2}\times \sqrt{2})R45^{\circ}$ ordered.
This leads to a natural explanation to the surface structure of
Ba122 cleaved at low temperature measured by Pan and
coworkers\cite{shpan}. Furthermore, we predict a switchable
$(\sqrt{2}\times \sqrt{2})R45^{\circ}$ STM pattern upon varying an
applied bias, which can be a hallmark for testing whether the
As-terminated or the Ba-terminated is. This theoretical prediction
can be further verified by annealing an As-terminated surface to
high temperatures so that the $A$ atoms can diffuse to form an
$A$-terminated surface. For the As-terminated Sr(Ca)122 surface,
there is no buckling, dimerization, or any other reconstruction.
Thus the STM-observed coexistence of $(\sqrt{2}\times
\sqrt{2})R45^{\circ}$ and $(1\times 2)$ patterns on Sr122
surface\cite{zqwang} should be attributed to the Sr-termination. For
Ca122, we find that the Ca-terminated surface is $(1\times 2)$
ordered which coexists partly with a $(\sqrt{2}\times
\sqrt{2})R45^{\circ}$ order. Finally, among these surface
structures, there are essential surface states induced crossing or
nearby the Fermi energy within the bulk energy band gap regions,
which will affect those surface sensitive experimental probing.


This work is partially supported by National Natural Science
Foundation of China and by National Program for Basic Research of
MOST, China. We would like to thank Prof. S. Pan for helpful
discussions.

\end{document}